\pdfoutput=1
\documentclass[12pt,a4paper]{article}

\usepackage{ifthen} 
\newboolean{pdflatex}
\setboolean{pdflatex}{true} 

\newboolean{articletitles}
\setboolean{articletitles}{true} 

\newboolean{uprightparticles}
\setboolean{uprightparticles}{false} 

\def\paperauthors{LHCb collaboration} 
\def\paperasciititle{Observation of exotic Jpsi-phi resonant structure in diffractive processes in proton-proton collisions} 
\def\papertitle{Observation of exotic ${J\mskip -3mu/\mskip -2mu\psi}\phi$ resonant structure in diffractive processes in proton-proton collisions} 
\def\paperkeywords{{High Energy Physics}, {LHCb}} 
\def\papercopyright{\the\year\ CERN for the benefit of the LHCb collaboration} 
\def\paperlicence{CC BY 4.0 licence}
\def\paperlicenceurl{https://creativecommons.org/licenses/by/4.0/}

\usepackage[top=1in, bottom=1.25in, left=1in, right=1in]{geometry}

\usepackage{microtype}
\usepackage{lineno}  
\usepackage{xspace} 
\usepackage{caption} 

\usepackage{graphicx}  
\usepackage{color}
\usepackage{colortbl}
\graphicspath{{./figs/}} 

\usepackage{amsmath} 
\usepackage{amssymb}
\usepackage{amsfonts}
\usepackage{upgreek} 

\newcommand*\patchAmsMathEnvironmentForLineno[1]{%
\expandafter\let\csname old#1\expandafter\endcsname\csname #1\endcsname
\expandafter\let\csname oldend#1\expandafter\endcsname\csname
end#1\endcsname
 \renewenvironment{#1}%
   {\linenomath\csname old#1\endcsname}%
   {\csname oldend#1\endcsname\endlinenomath}%
}
\newcommand*\patchBothAmsMathEnvironmentsForLineno[1]{%
  \patchAmsMathEnvironmentForLineno{#1}%
  \patchAmsMathEnvironmentForLineno{#1*}%
}
\AtBeginDocument{%
\patchBothAmsMathEnvironmentsForLineno{equation}%
\patchBothAmsMathEnvironmentsForLineno{align}%
\patchBothAmsMathEnvironmentsForLineno{flalign}%
\patchBothAmsMathEnvironmentsForLineno{alignat}%
\patchBothAmsMathEnvironmentsForLineno{gather}%
\patchBothAmsMathEnvironmentsForLineno{multline}%
\patchBothAmsMathEnvironmentsForLineno{eqnarray}%
}

\usepackage{hyperxmp}

\usepackage[pdftex,
            pdfauthor={\paperauthors},
            pdftitle={\paperasciititle},
            pdfkeywords={\paperkeywords},
            pdfcopyright={Copyright (C) \papercopyright},
            pdflicenseurl={\paperlicenceurl}]{hyperref}

\usepackage[colorinlistoftodos,textsize=scriptsize]{todonotes}

\usepackage[bottom,flushmargin,hang,multiple]{footmisc}

\usepackage[all]{hypcap} 

\usepackage{xspace} 
\usepackage{upgreek}

\def\Purity {\ensuremath{(93.0 \pm 0.5)\%}\xspace}

\def\xfouronefourcs{\ensuremath{(0.80 {\pm 0.15} \, {\pm 0.28}})\pb}
\def\xfourtwosevenfourcs{\ensuremath{(0.73 {\pm 0.08} \, {\pm 0.17}})\pb}
\def\xfourfivecs{\ensuremath{(0.42 {^{+0.09}_{-0.08}} \, {\pm 0.06}})\pb}
\def\xfoursevencs{\ensuremath{(0.14 {^{+0.07}_{-0.06}} \, {\pm 0.06}})\pb}
\def\NRcs{\ensuremath{(0.43 {^{+0.24}_{-0.18}} \, {\pm 0.20}})\pb}
\def\Totalcs{\ensuremath{(2.52 \pm {0.08} \pm {0.12} \pm {0.05})\pb}}

\def\Mfourtwosevenfour{\ensuremath{4298 {\pm 6} \, {\pm 9}}}
\def\Wfourtwosevenfour{\ensuremath{92 {^{+22}_{-18}} \, {\pm 57}}}
\def\Mfourfive{\ensuremath{4512.5 {^{+6.0}_{-6.2}} \, {\pm 3.0}}}
\def\Wfourfive{\ensuremath{65 {^{+20}_{-16}} \, {\pm 32}}}

\def\XtoJPsiPhi {\ensuremath{X\to\jpsi\Pphi}\xspace}
\def\BtoJPsiPhiK {\ensuremath{\Bu\to\jpsi\Pphi\Kp}\xspace}
\def\jpsiphi  {\ensuremath{\jpsi\phi}\xspace}
\def\jpsiphidecay {\decay{\jpsi(}{\mup\mun})\decay{\phi(}{\Kp\Km})\xspace}
\def\xfouronefour  {{\ensuremath{\chi_{c1}(4140)}}\xspace}
\def\xfourtwosevenfour  {{\ensuremath{\chi_{c1}(4274)}}\xspace}
\def\xfourfive  {{\ensuremath{\chi_{c0}(4500)}}\xspace}
\def\xfoursixeight {{\ensuremath{\chi_{c1}(4685)}}\xspace}
\def\xfourseven  {{\ensuremath{\chi_{c0}(4700)}}\xspace}

\def\nr  {{\ensuremath{\mathrm{NR}}}\xspace}

\def\mjpsiphi {\ensuremath{M_{\jpsiphi}}\xspace}

\def\lhcb   {\mbox{LHCb}\xspace}

\def\velo   {VELO\xspace}

\def\herschel {\mbox{\textsc{HeRSCheL}}\xspace}
\def\spd    {SPD\xspace}

\def\MagUp {\mbox{\em Mag\kern -0.05em Up}\xspace}

\ifthenelse{\boolean{uprightparticles}}%
{

 \def\Peta        {\ensuremath{\upeta}\xspace}

 \def\Pmu         {\ensuremath{\upmu}\xspace}

 \def\Pphi        {\ensuremath{\upphi}\xspace}

 \def\Ppsi        {\ensuremath{\uppsi}\xspace}

 \def\PDelta      {\ensuremath{\Delta}\xspace}                 
 \def\PXi         {\ensuremath{\Xi}\xspace}                 
 \def\PLambda     {\ensuremath{\Lambda}\xspace}                 
 \def\PSigma      {\ensuremath{\Sigma}\xspace}                 
 \def\POmega      {\ensuremath{\Omega}\xspace}                 
 \def\PUpsilon    {\ensuremath{\Upsilon}\xspace}

 \def\PB      {\ensuremath{\mathrm{B}}\xspace}                 
                  
 \def\PD      {\ensuremath{\mathrm{D}}\xspace}

 \def\PJ      {\ensuremath{\mathrm{J}}\xspace}                 
 \def\PK      {\ensuremath{\mathrm{K}}\xspace}

 \def\Pi      {\ensuremath{\mathrm{i}}\xspace}

 \def\Pp      {\ensuremath{\mathrm{p}}\xspace}

 \def\Ps      {\ensuremath{\mathrm{s}}\xspace}

 \def\thebaroffset{0.0em}
}
{

 \def\Peta        {\ensuremath{\eta}\xspace}

 \def\Pmu         {\ensuremath{\mu}\xspace}

 \def\Pphi        {\ensuremath{\phi}\xspace}

 \def\Ppsi        {\ensuremath{\psi}\xspace}                 
                  
 \mathchardef\PDelta="7101
 \mathchardef\PXi="7104
 \mathchardef\PLambda="7103
 \mathchardef\PSigma="7106
 \mathchardef\POmega="710A
 \mathchardef\PUpsilon="7107
                  
 \def\PB      {\ensuremath{B}\xspace}                 
                  
 \def\PD      {\ensuremath{D}\xspace}

 \def\PJ      {\ensuremath{J}\xspace}                 
 \def\PK      {\ensuremath{K}\xspace}

 \def\Pi      {\ensuremath{i}\xspace}

 \def\Pp      {\ensuremath{p}\xspace}

 \def\Ps      {\ensuremath{s}\xspace}

 \def\thebaroffset{0.18em}
}
\newcommand{\offsetoverline}[2][\thebaroffset]{\kern #1\overline{\kern -#1 #2}}%

\makeatletter
\ifcase \@ptsize \relax
  \newcommand{\miniscule}{\@setfontsize\miniscule{4}{5}}
\or
  \newcommand{\miniscule}{\@setfontsize\miniscule{5}{6}}
\or
  \newcommand{\miniscule}{\@setfontsize\miniscule{5}{6}}
\fi
\makeatother

\DeclareRobustCommand{\optbar}[1]{\shortstack{{\miniscule (\rule[.5ex]{1.25em}{.18mm})}
  \\ [-.7ex] $#1$}}

\def\mup        {{\ensuremath{\Pmu^+}}\xspace}
\def\mun        {{\ensuremath{\Pmu^-}}\xspace} 

\def\mumu       {{\ensuremath{\Pmu^+\Pmu^-}}\xspace}

\def\squark    {{\ensuremath{\Ps}}\xspace}

\def\kaon    {{\ensuremath{\PK}}\xspace}

\def\KorKbar {\kern \thebaroffset\optbar{\kern -\thebaroffset \PK}{}\xspace}

\def\Kp      {{\ensuremath{\kaon^+}}\xspace}
\def\Km      {{\ensuremath{\kaon^-}}\xspace}

\def\D       {{\ensuremath{\PD}}\xspace}

\def\DorDbar {\kern \thebaroffset\optbar{\kern -\thebaroffset \PD}\xspace}

\def\Dp      {{\ensuremath{\D^+}}\xspace}
\def\Dm      {{\ensuremath{\D^-}}\xspace}

\def\DpDm    {\ensuremath{\Dp {\kern -0.16em \Dm}}\xspace}

\def\B       {{\ensuremath{\PB}}\xspace}

\def\BorBbar {\kern \thebaroffset\optbar{\kern -\thebaroffset \PB}\xspace}

\def\Bd      {{\ensuremath{\B^0}}\xspace}

\def\BdorBdbar {\kern \thebaroffset\optbar{\kern -\thebaroffset \Bd}\xspace}
\def\Bu      {{\ensuremath{\B^+}}\xspace}

\def\Bs      {{\ensuremath{\B^0_\squark}}\xspace}

\def\BsorBsbar {\kern \thebaroffset\optbar{\kern -\thebaroffset \Bs}\xspace}

\def\jpsi     {{\ensuremath{{\PJ\mskip -3mu/\mskip -2mu\Ppsi}}}\xspace}
\def\psitwos  {{\ensuremath{\Ppsi{(2S)}}}\xspace}

\def\Y#1S{\ensuremath{\PUpsilon{(#1S)}}\xspace}

\def\proton      {{\ensuremath{\Pp}}\xspace}

\def\LorLbar     {\kern \thebaroffset\optbar{\kern -\thebaroffset \PLambda}\xspace}

\def\BF         {{\ensuremath{\mathcal{B}}}\xspace}

\newcommand{\decay}[2]{\ensuremath{#1\!\to #2}\xspace} 

\def\to                 {\ensuremath{\rightarrow}\xspace}

\def\AT#1     {\ensuremath{A_{\mathrm{T}}^{#1}}\xspace}           

\def\C#1      {\ensuremath{\mathcal{C}_{#1}}\xspace}                       
\def\Cp#1     {\ensuremath{\mathcal{C}_{#1}^{'}}\xspace}                    
\def\Ceff#1   {\ensuremath{\mathcal{C}_{#1}^{\mathrm{(eff)}}}\xspace}        
\def\Cpeff#1  {\ensuremath{\mathcal{C}_{#1}^{'\mathrm{(eff)}}}\xspace}       
\def\Ope#1    {\ensuremath{\mathcal{O}_{#1}}\xspace}                       
\def\Opep#1   {\ensuremath{\mathcal{O}_{#1}^{'}}\xspace}                    

\newcommand{\aunit}[1]{\ensuremath{\text{\,#1}}}       

\newcommand{\tev}{\aunit{Te\kern -0.1em V}\xspace}
\newcommand{\gev}{\aunit{Ge\kern -0.1em V}\xspace}
\newcommand{\mev}{\aunit{Me\kern -0.1em V}\xspace}
\newcommand{\kev}{\aunit{ke\kern -0.1em V}\xspace}
\newcommand{\ev}{\aunit{e\kern -0.1em V}\xspace}
 
\newcommand{\mevc}{\ensuremath{\aunit{Me\kern -0.1em V\!/}c}\xspace}
\newcommand{\gevc}{\ensuremath{\aunit{Ge\kern -0.1em V\!/}c}\xspace}
\newcommand{\mevcc}{\ensuremath{\aunit{Me\kern -0.1em V\!/}c^2}\xspace}
\newcommand{\gevcc}{\ensuremath{\aunit{Ge\kern -0.1em V\!/}c^2}\xspace}

\def\pb {\aunit{pb}\xspace}

\def\fb   {\ensuremath{\aunit{fb}}\xspace}
\def\invfb   {\ensuremath{\fb^{-1}}\xspace}

\def\gsim{{~\raise.15em\hbox{$>$}\kern-.85em
          \lower.35em\hbox{$\sim$}~}\xspace}
\def\lsim{{~\raise.15em\hbox{$<$}\kern-.85em
          \lower.35em\hbox{$\sim$}~}\xspace}

\def\sqs   {\ensuremath{\protect\sqrt{s}}\xspace}

\def\pt         {\ensuremath{p_{\mathrm{T}}}\xspace}

\newcommand{\lum} {\ensuremath{\mathcal{L}}\xspace}

\def\geant      {\mbox{\textsc{Geant4}}\xspace}
\def\superchic  {\mbox{\textsc{SuperChic3}}\xspace}
\def\superchict {\mbox{\textsc{SuperChic2}}\xspace}

\def\photos     {\mbox{\textsc{Photos}}\xspace}

\def\tell1  {TELL1\xspace}
\def\ukl1   {UKL1\xspace}

\newcommand{\ie}{\mbox{\itshape i.e.}\xspace}

\newcommand{\lhcborcid}[1]{\href{https://orcid.org/#1}{\hspace*{0.1em}\raisebox{-0.45ex}{\includegraphics[width=1em]{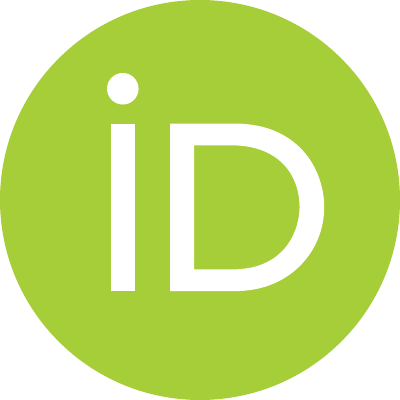}}}}

\usepackage{cite} 
\usepackage{mciteplus}

\usepackage{longtable} 
\usepackage{multirow}

\begin{document}

\renewcommand{\thefootnote}{\fnsymbol{footnote}}
\setcounter{footnote}{1}


\begin{titlepage}
\pagenumbering{roman}

\vspace*{-1.5cm}
\centerline{\large EUROPEAN ORGANIZATION FOR NUCLEAR RESEARCH (CERN)}
\vspace*{1.5cm}
\noindent
\begin{tabular*}{\linewidth}{lc@{\extracolsep{\fill}}r@{\extracolsep{0pt}}}
\ifthenelse{\boolean{pdflatex}}
{\vspace*{-1.5cm}\mbox{\!\!\!\includegraphics[width=.14\textwidth]{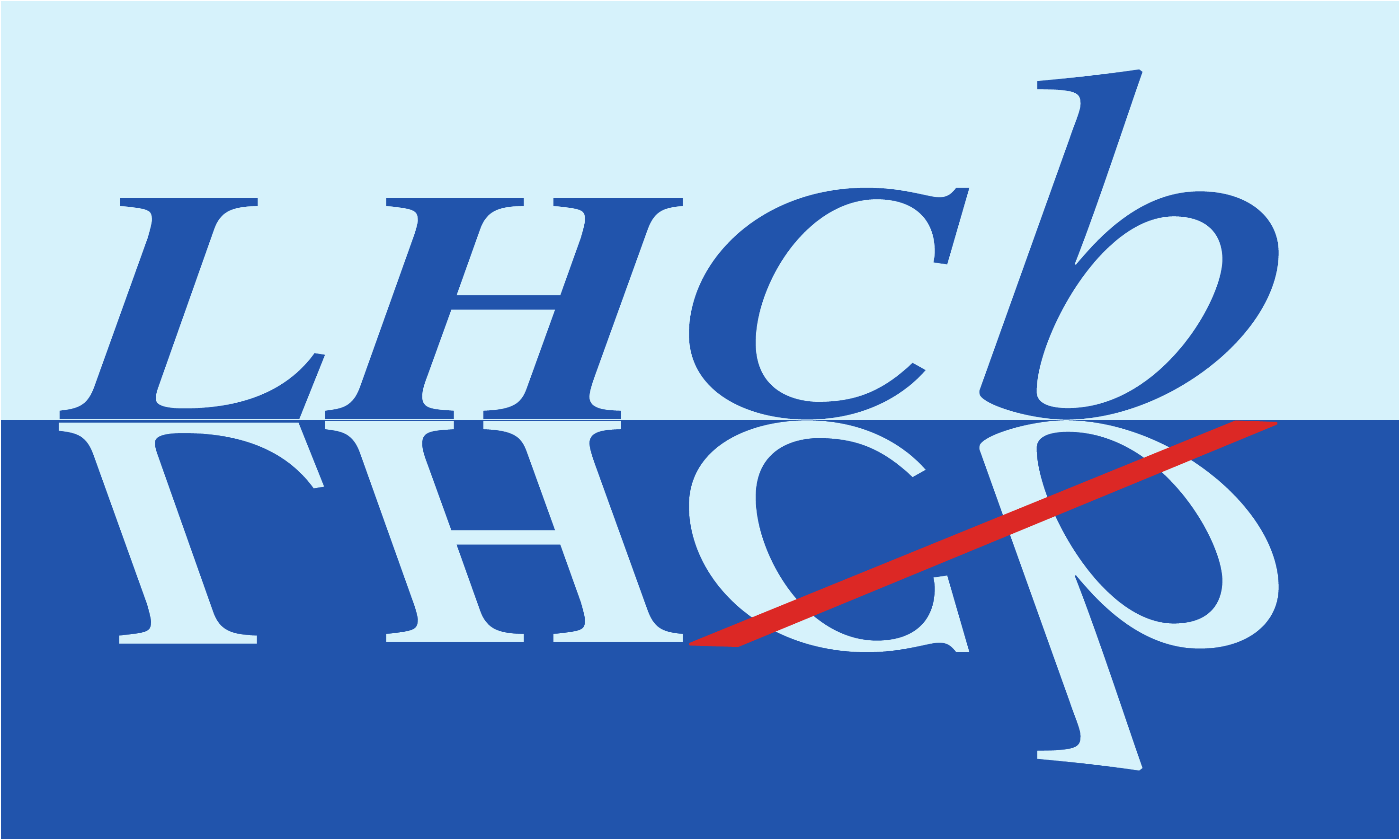}} & &}%
{\vspace*{-1.2cm}\mbox{\!\!\!\includegraphics[width=.12\textwidth]{figs/lhcb-logo.eps}} & &}%
\\
 & & CERN-EP-2024-149 \\  
 & & LHCb-PAPER-2023-043 \\  
 & & January 24, 2025 \\ 
\end{tabular*}

\vspace*{4.0cm}

{\normalfont\bfseries\boldmath\huge
\begin{center}
  \papertitle 
\end{center}
}

\vspace*{2.0cm}

\begin{center}
\paperauthors\footnote{Authors are listed at the end of this paper.}
\end{center}

\vspace{\fill}

\begin{abstract}
  \noindent

The first study of \mbox{$J\mskip -3mu/\mskip -2mu\psi\phi$} production in diffractive processes in proton-proton collisions is presented.
The study is based on an LHCb dataset recorded at centre-of-mass energy of \mbox{$13\,\mathrm{TeV}$}, corresponding to an integrated luminosity of $5\,\mathrm{fb}^{-1}$.
The data disfavour a nonresonant \mbox{$J\mskip -3mu/\mskip -2mu\psi\phi$} production but are consistent with a resonant model including several resonant states observed previously only in \mbox{$\B^+\to J\mskip -3mu/\mskip -2mu\psi\phi K^+$} decays.
The $\chi_{c0}(4500)$ state is observed with a significance over $6\,\sigma$ and the $\chi_{c1}(4274)$ is confirmed with a significance of more than $4\,\sigma$.
\end{abstract}

\vspace*{2.0cm}

\begin{center}
  Published in Phys. Rev. Lett. 134 (2025) 031902
\end{center}

\vspace{\fill}

{\footnotesize 
\centerline{\copyright~\papercopyright. \href{\paperlicenceurl}{\paperlicence}.}}
\vspace*{2mm}

\end{titlepage}


\newpage
\setcounter{page}{2}
\mbox{~}
%
%
%
%


\renewcommand{\thefootnote}{\arabic{footnote}}
\setcounter{footnote}{0}

\cleardoublepage


\pagestyle{plain} 
\setcounter{page}{1}
\pagenumbering{arabic}


Studies of exclusive hadronic processes in proton-proton (\proton\proton) collisions have been proposed to clarify poorly understood phenomena associated with the asymptotic high-energy behaviour of quantum chromodynamics (QCD)~\cite{Brodsky:1974vy}.
In particular, it remains theoretically challenging to describe the production of complex states, such as vector-meson pairs, in central diffractive collisions with large rapidity gaps. 
Photon-induced or pomeron-induced processes have been proposed to understand the QCD dynamics at high energies~\cite{Bertulani:1987tz} including the production of exotic states and double vector-meson production~\cite{Harland-Lang_2015,Goncalves:2016ybl,Pan:2019sxp, Lebiedowicz:2016zka,Moreira:2016ciu,Goncalves:2018hiw,Xie:2018ict}.
The discovery of exotic QCD states has motivated an extensive theoretical and experimental effort to understand their properties~\cite{Brambilla:2019esw}.
The nature of exotic hadrons has been studied mainly in inclusive production or in exclusive beauty hadron decays, in terms of properties such as mass, spin and decay width.
Resonant structures in the \jpsiphi mass spectrum have previously been observed in amplitude analyses of \BtoJPsiPhiK decays~\cite{LHCb-PAPER-2016-018,LHCb-PAPER-2016-019,LHCb-PAPER-2020-044}.
Five of those exotic candidates, \xfouronefour, \xfourtwosevenfour and \xfoursixeight with quantum numbers $J^{PC} = 1^{++}$ and \xfourfive and \xfourseven with $J^{PC} = 0^{++}$, can be produced in photon-photon or pomeron-pomeron processes, with the latter expected to dominate in \proton\proton collisions.
Searches for the production of these exotic-hadron candidates in such processes can help elucidate their nature and distinguish between compact tetraquarks and molecular states~\cite{Brodsky:2015wza,Cerri:2018ypt,PhysRevD.104.114029,Esposito}.

In this Letter, the \jpsiphidecay production cross-section is measured for the first time in events with no additional detected activity, in which photon- and pomeron-induced processes are expected to be dominant.
An analysis of the \jpsiphi invariant-mass distribution is also performed and production cross-sections are determined for each of the five resonant states and a nonresonant (\nr) component.
The data used correspond to an integrated luminosity of $5\invfb$ collected in $pp$ collisions at $\sqs=13\tev$ between 2016 and 2018 with the \lhcb detector.

The \lhcb detector is a single-arm forward spectrometer covering the pseudorapidity range $2<\eta<5$, described in detail in Refs.~\cite{LHCb-DP-2008-001,LHCb-DP-2014-002}.
The \herschel detector~\cite{LHCb-DP-2016-003} extends the \lhcb coverage to the pseudorapidity range $5<|\eta|<10$, increasing sensitivity to proton dissociation background.
A simulated dataset of $\proton\proton\to\proton+\jpsiphi+\proton$ events is created using the \superchic package~\cite{superchic3,Harland-Lang_2015}, by generating $\proton\jpsi\jpsi\proton$ events and replacing one of the \jpsi states with a \Pphi resonance.
Final-state radiation is simulated using the \photos package \cite{davidson2015photos}.
The interaction of the generated particles with the detector material, and the detector response, are simulated using the \geant toolkit~\cite{Agostinelli:2002hh,Allison:2006ve} as described in Ref.~\cite{LHCb-PROC-2011-006}.
Samples of photoproduced \jpsi mesons, generated with \superchict~\cite{superchic2}, are used to constrain the \jpsi mass model.

The \XtoJPsiPhi candidate decays, where $X$ indicates a hypothetical resonance, are selected online by a two-stage trigger optimised to identify low-multiplicity events with at least one muon.
The first stage requires fewer than 20 deposits in the scintillating-pad detector (\spd) of the calorimeter system and at least one muon candidate with a transverse momentum (\pt) greater than $200\mev$.\footnote{Natural units with $\hbar=c=1$ are used throughout.}
The second stage applies a full event reconstruction and requires fewer than eight tracks with pseudorapidity $2<\eta<5$ (forward tracks), no tracks with pseudorapidity $-3.5<\eta<-1.5$ (backward tracks) and at least one track identified as a muon with \pt above $400\mev$.

Additional geometric and kinematic requirements are applied to events selected by the trigger system.
Particle identification information is used to select two kaon and two muon candidates, which must also have \pt above $200\mev$ and be in the region $2<\eta< 5$.
At least one muon candidate is required to be responsible for the event passing the hardware and software stages of the trigger.
Events with additional activity, \ie, with additional tracks reconstructed in the vertex detector (\velo tracks), are vetoed.
The dimuon (dikaon) invariant mass is required to be in the range \mbox{$3036 < M_{\mu\mu} < 3156\mev$} (\mbox{$1005 < M_{\PK\PK} < 1035\mev$}).
The \jpsiphi invariant mass is required to be less than $6000\mev$.
After the selection requirements, 989 \jpsiphi candidates are retained.

No requirements are applied to discriminate events in which the outgoing protons stay intact from those where one or both of the protons dissociate.
Instead, a variable that quantifies activity above noise in the \herschel detector is used to estimate the fraction for each of these cases.
The efficiency of such a classification is determined using the same method described in Ref.~\cite{LHCb-PAPER-2018-011} and it is found that at least $69\%$ of the \jpsiphi candidates are produced in events in which one or both protons dissociate.

The requirement of no additional activity results in the rejection of events that are accompanied by additional visible \proton\proton interactions in the same bunch crossing.
Correspondingly, a single-interaction integrated luminosity ($\lum_{\mathrm{eff}}$) must be determined to evaluate the cross-section.
The number of visible $\proton\proton$ interactions per bunch crossing is assumed to follow a Poisson distribution.
The mean is determined using the fraction of beam crossings with no visible activity, \ie, with no reconstructed \velo tracks.
The effective single-interaction integrated luminosity is found to be about $1.75\invfb$.
The integrated cross-section is then calculated as
\begin{equation}
    \sigma = \frac{\mathcal{P} \cdot N}{\varepsilon \cdot \lum_{\mathrm{eff}}} \mathrm{,}
\end{equation}
where $N$ is the total event yield (or individual component yield), $\mathcal{P}$ is the sample purity and $\varepsilon$ is the total efficiency, which includes reconstruction and selection efficiencies.
Note that 46 of the selected events were collected during moments in which the luminosity could not be reliably measured.
Therefore, these events are excluded from the cross-section calculation.

\begin{figure}[h]
    \centering
    \includegraphics[width=0.7\linewidth]{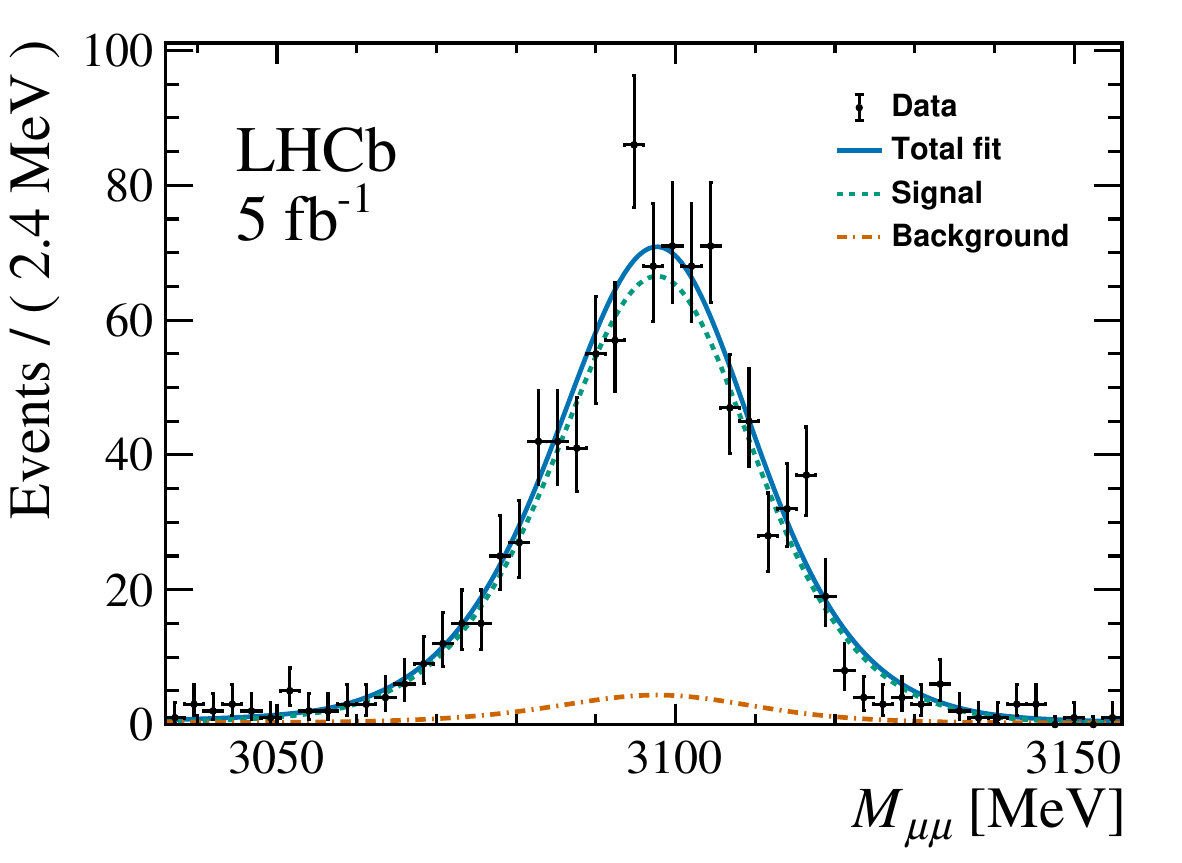}
    \includegraphics[width=0.7\linewidth]{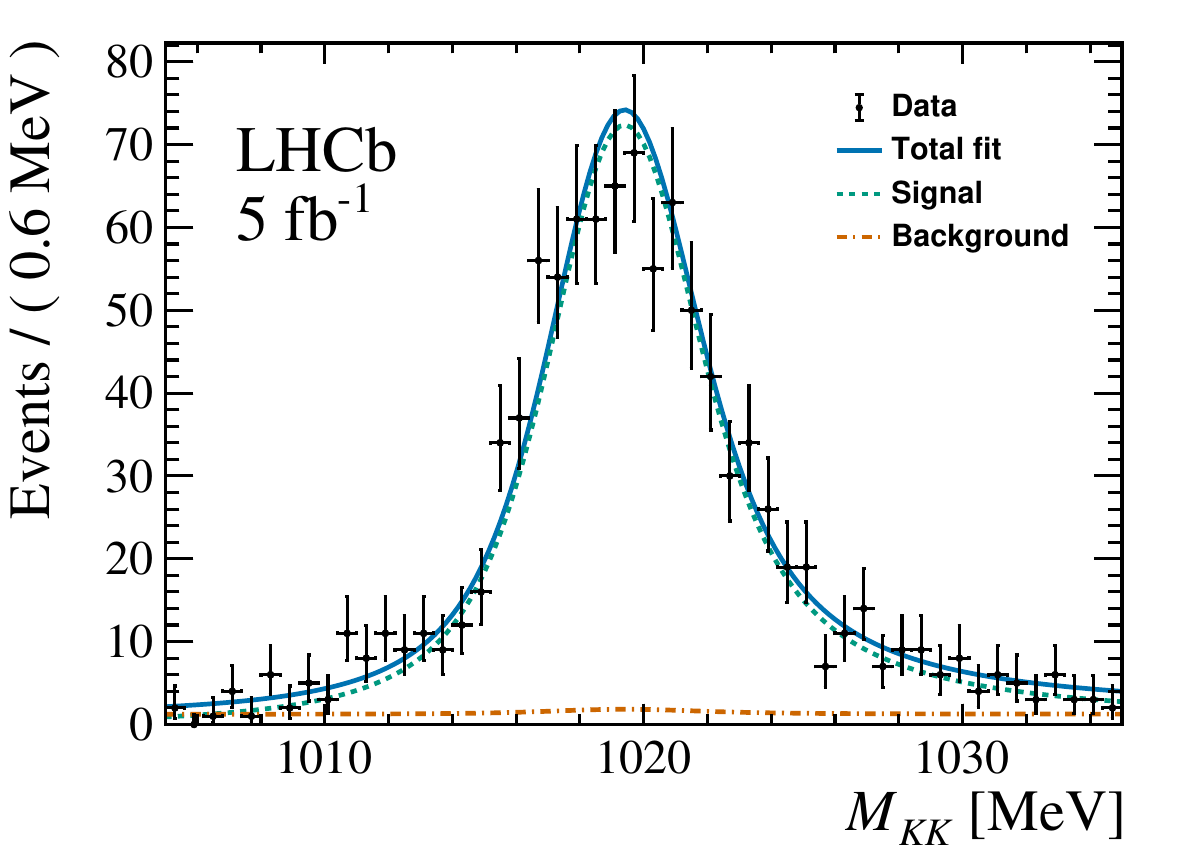}
    \caption{\small (Top) Dimuon and (bottom) dikaon invariant-mass distributions within the invariant-mass selection windows. The projections of the two-dimensional unbinned fit are overlaid.
    The signal component consists of $J\mskip -3mu/\mskip -2mu\psi\phi$ candidates.
    The peaking behaviour of the background arises when one of the pairs ($\mu\mu$ or $KK$) is background.}
    \label{fig:jpsixphi_mass_fits_selection}
\end{figure}

In order to estimate the sample purity, \ie the fraction of events with resonant dimuon and dikaon production, a two-dimensional unbinned extended maximum-likelihood fit is performed on the dimuon and dikaon invariant-mass distributions.
For this, the invariant-mass window requirements are relaxed to \mbox{$1500 < M_{\mu\mu} < 4500\mev$} and \mbox{$990 < M_{KK} < 1190\mev$}.
The \jpsi contribution is modelled by a double-sided Crystal Ball function~\cite{Skwarnicki:1986xj} with tail parameters determined from simulation, while the \psitwos is modelled by a Gaussian function and the nonresonant $\mumu$ background with an exponential function.
The \Pphi contribution is modelled by a relativistic Breit--Wigner function convolved with a Gaussian function, while the nonresonant $\Kp\Km$ background is described by a uniform distribution.
The sample purity is found to be \Purity.
The efficiency of the mass-window requirement is estimated as the fraction of events in the mass window with respect to the relaxed mass range.
The dimuon and dikaon invariant-mass distributions, along with the projections of the fit result, are shown in Fig.~\ref{fig:jpsixphi_mass_fits_selection}.

The efficiency of the \spd requirement is estimated using a combination of simulation and data samples.
The simulated \spd multiplicity from signal events is convolved with the distribution of spillover \spd hits from the previous bunch crossing in data, obtained from events with no visible activity, and the fraction of events passing the requirement is determined.
The efficiency of vetoing additional \velo tracks is studied using simulation.

The reconstruction efficiencies are obtained from simulation, and correspond to the fraction of charged particles produced in the \lhcb angular acceptance that are reconstructed as tracks.
The simulated sample is also used to determine the muon-system acceptance, which is calculated as the fraction of reconstructed muons that cross the \lhcb muon system.
The muon identification efficiency is obtained with a data-driven method using a sample of $\jpsi\to\mumu$ decays in low multiplicity events.
The same data sample is used to calculate the muon trigger efficiency.
Since the muon trigger requirements include muon identification, this efficiency is calculated for identified muons that satisfy the muon trigger requirements.
The efficiency for kaon identification is obtained with a data-driven method using a sample of $\Pphi\to\Kp\Km$ decays in low multiplicity events.
The muon and kaon detection efficiencies, which are computed as functions of \pt and \Peta, are assumed to factorise.
The overall selection efficiency is found to be $(19.90 \pm 0.32)\%$, where the uncertainty is statistical.

The \jpsiphi invariant-mass spectrum is described by a model consisting of five resonances, namely \xfouronefour, \xfourtwosevenfour, \xfourfive, \xfoursixeight and \xfourseven, and a nonresonant component.
The resonances considered are the ones observed in Refs.~\cite{LHCb-PAPER-2016-018,LHCb-PAPER-2016-019,LHCb-PAPER-2020-044} with quantum numbers compatible with diffractive photon-induced or pomeron-induced production processes.
The resonant components are described by relativistic Breit--Wigner distributions convolved with Gaussian functions to account for detector resolution effects.
The resolutions are estimated from simulation.
Due to the small sample size, it is assumed that there is no interference between the fit components, and the mass and width of the less significant resonances (\xfouronefour, \xfoursixeight and \xfourseven) are fixed to the values observed in Ref.~\cite{LHCb-PAPER-2020-044}.
Systematic uncertainties are estimated to account for the possible effects of interference in the fit parameters.
Moreover, only the combined $\xfoursixeight+\xfourseven$ yield is estimated since it is not possible to resolve the \xfoursixeight and \xfourseven states.
The nonresonant component is modelled by an exponential distribution modified with a parametrised factor to account for the \jpsiphi mass threshold,
\begin{equation}
\label{eq:th_factor}
    f_{\nr}=\bigg[1 - \mathrm{exp}\Big(a\cdot \Big(\frac{M_{\jpsiphi}}{m_{T}} - 1\Big)\Big)\bigg] \cdot \mathrm{exp}\Big(b \cdot \frac{M_{\jpsiphi}}{m_{T}}\Big) \mathrm{,}
\end{equation}
where $m_T$ is the \jpsiphi mass threshold, $a$ is the coefficient representing the strength of the threshold effect and $b$ is the slope of the exponential distribution.
The nonresonant component is also convolved with a Gaussian function to account for detector resolution.

\begin{figure}[h]
    \centering
    \includegraphics[width=\linewidth]{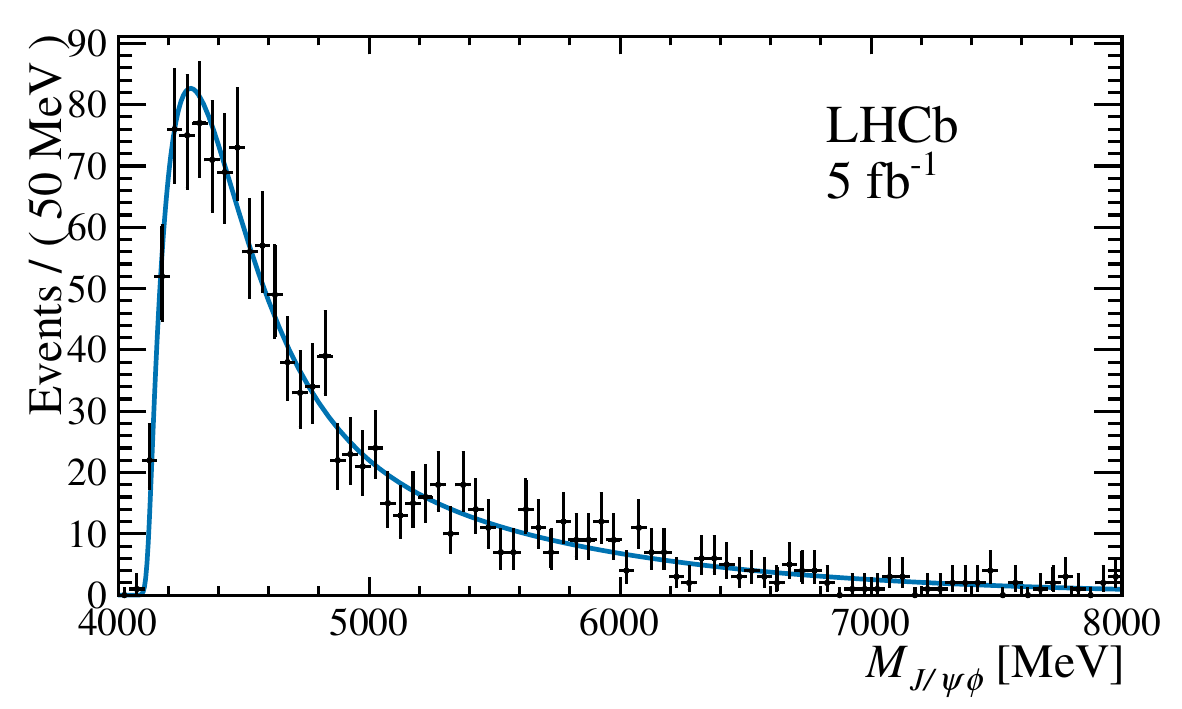}
    \caption{\small Invariant-mass distribution of \mbox{$J\mskip -3mu/\mskip -2mu\psi\phi$} candidates for the sideband sample composed of events with more than four reconstructed VELO tracks.
    Note that all other selection requirements are applied, including the $J\mskip -3mu/\mskip -2mu\psi$ and $\phi$ mass-window requirements.
    The fit results are shown as a solid line.}
    \label{fig:nr_only_fit}
\end{figure}

\begin{figure}[h]
    \centering
    \includegraphics[width=\linewidth]{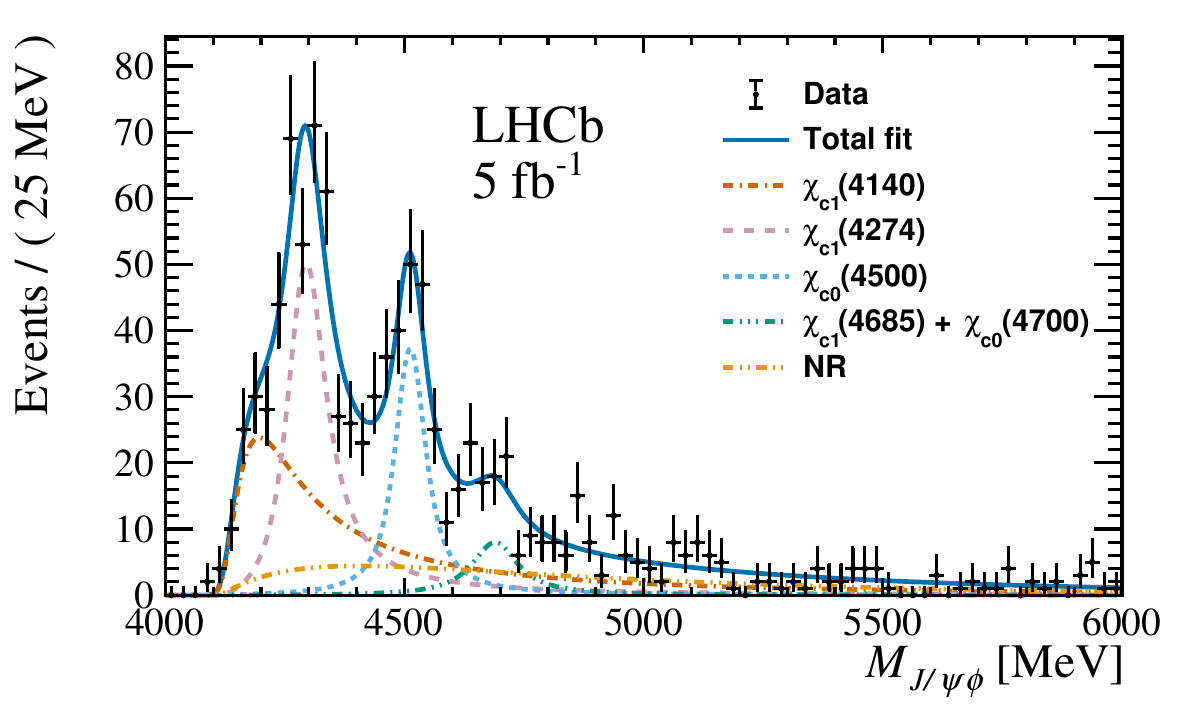}
    \caption{\small Invariant-mass distribution of \mbox{$J\mskip -3mu/\mskip -2mu\psi(\!\to\mu^+\mu^-)\phi(\!\to K^+K^-)$} candidates in the signal sample after selection. The fit components are shown as dashed lines.}
    \label{fig:full_model_fit}
\end{figure}

The parameter $a$ is determined separately using a sideband sample, which is obtained by applying all selection requirements, except for an inverted offline multiplicity requirement of more than four \velo tracks.
Figure~\ref{fig:nr_only_fit} shows the result of a maximum-likelihood fit performed on the resulting distribution using a model that consists of a sum of two exponential functions modified by the turn-on factor described in Eq.~\ref{eq:th_factor} and convolved with the same resolution function.
This distribution shows no clear mass structure, however when imposing the veto on additional \velo tracks, a resonant structure appears, which indicates the presence of a photon-induced or pomeron-induced mechanism.
The resulting \jpsiphi invariant-mass distribution for signal candidates is shown in Fig.~\ref{fig:full_model_fit} overlaid with the results of an extended maximum-likelihood fit of the signal model.

The sources of systematic uncertainties considered for the cross-section measurement are summarised below.
The integrated luminosity is determined with a precision of $2\%$~\cite{LHCb-PAPER-2014-047}.
The uncertainty on the \spd requirement efficiency, which accounts for possible mismodeling of the \spd distribution, is estimated to be $1.3\%$ by shifting the model until it no longer describes data.
The systematic uncertainty due to the size of the simulated sample used to estimate the efficiency for vetoing additional \velo tracks is found to be smaller than $0.1\%$.
The muon trigger, muon identification and kaon identification efficiencies uncertainties due to the limited data samples are found to be below $0.2\%$.

Systematic uncertainties related to the sample purity and efficiency of invariant-mass requirements were determined from the two-dimensional fit and are $0.5\%$ and $0.04\%$, respectively.
These uncertainties are combined via linear propagation using the correlation matrix from the fit.
Furthermore, an alternative parametrisation of the background in the dikaon invariant-mass spectrum is evaluated, which consists of a second-order polynomial.
This causes a $0.4\%$ variation in the sample purity, which is added to the corresponding systematic uncertainty, resulting in a overall uncertainty of $0.7\%$.
An additional systematic uncertainty is considered for the cross-section of the individual \jpsiphi components due to the possible dependence of the purity on the \jpsiphi invariant mass.
The two-dimensional fit to the dimuon and dikaon invariant-mass distributions shows that the background is dominated by \jpsi\Kp\Km candidates.
A fit to the \jpsi\Kp\Km invariant-mass distribution is performed using events selected with \mbox{$1050<M_{KK}<1200\mev$}.
The change in the relative contribution of the resonant and nonresonant components is found to be $0.5\%$ and $2.4\%$, respectively, and is taken as systematic uncertainty.

The systematic uncertainties on the kaon and muon reconstruction efficiencies and muon system acceptance due to the limited simulation sample size are found to be negligible.
To account for the systematic uncertainty due to the discrepancies between data and simulation in the tracking efficiency, a $0.4\%$ systematic uncertainty per track and an additional $1.4\%$ systematic uncertainty per hadron is included~\cite{LHCb-DP-2013-002}.
It is assumed that the $0.4\%$ uncertainty is fully correlated for the 4 tracks, as well as the $1.4\%$ factor for the two hadrons, which amounts to an extra $3.2\%$ systematic uncertainty.

The factorisation hypothesis employed for the efficiencies is tested using simulation.
Simulated \jpsiphidecay events are selected by requiring that the kaon and muon tracks satisfy the nominal \pt and \Peta selection requirements, that the kaons and muons are reconstructed, and that the muons are within the muon system acceptance.
The resulting \mjpsiphi distribution is then corrected by the reconstruction efficiencies and muon system acceptance.
The corrected distribution is compared to the mass distribution without the reconstruction requirements after weighting the simulated events to have the same two-dimensional rapidity distribution, \mbox{$y_\jpsi \times y_\Pphi$}, as obtained in data.
It is found that the shapes of the distribution with no reconstruction requirements and the one corrected by the efficiencies are consistent.
However, the normalisation is underestimated by about $3\%$, which is included as a systematic uncertainty.
The individual systematic uncertainties are combined assuming no correlations, except where explicitly said otherwise, and the total relative systematic uncertainty of the total cross-section is $5.6\%$.

To assess the systematic uncertainty on the \jpsiphi invariant-mass fit parameters, a weighted fit is performed with each event-dependent correction varied within its uncertainty.
The threshold-effect strength parameter is varied within its uncertainties to study the impact of its precision.
Systematic uncertainties due to the \jpsiphi mass resolution determination are also considered, taking into account the precision of the Gaussian resolution and the choice of the resolution model.

To estimate the systematic uncertainties due to the fixed parameters in the model, the \jpsiphi invariant-mass fit is repeated with one mean and width pair allowed to vary at a time within the uncertainties quoted in Ref.\cite{LHCb-PAPER-2020-044}.
The resulting discrepancy on the remaining parameters with respect to the baseline fit is taken as a systematic uncertainty.
Systematic uncertainties due to unknown interference effects between the fit components are also considered.
These are estimated by repeating the \jpsiphi invariant-mass fits and allowing for interference with all possible phases between states with the same quantum numbers, including the nonresonant component.

The dominant sources of systematic uncertainties are summarised in Table~\ref{tab:sys} for the measured cross-sections, means and widths.
Moreover, the masses and widths obtained from the fit are compared to those in Ref.~\cite{LHCb-PAPER-2020-044} in Table~\ref{tab:comparison}.
The observed parameters are found to be in agreement with the previous \lhcb measurement, except for the \xfourfive mass, which is observed to have a higher value in this analysis.

\begin{table}[h]
    \begin{center}
        \caption{Summary of systematic uncertainties in the production cross-section ($\sigma$), mass ($M$) and width ($\Gamma$) of the measured resonant states. The contributions that affect the total uncertainty by less than 5\% are omitted. The two leftmost sources represent the contributions due to fixed parameters. The contribution due to the precision of the \jpsiphi invariant-mass resolution determination is indicated as $\Delta M$. The contributions due to unknown interference effects between $0^{++}$ states (including the nonresonant component) and between $1^{++}$ states are indicated as $\delta_{0++}$ and $\delta_{1++}$, respectively.}
        \begin{tabular}{l c c c c c c}
        \hline
        Source                                  & \xfouronefour & \xfourseven & $\Delta M$ & $\delta_{0++}$ & $\delta_{1++}$ & Total \\
        \hline
        $M_\xfourtwosevenfour\,[\!\mev]$        & $1.7$         & $0.2$       & $1.5$      & $^{+2.3}_{-0.7}$ & $^{+2.8}_{-4.3}$ & $^{+4}_{-5}$     \\
        $\Gamma_\xfourtwosevenfour\ \,[\!\mev]$ & $10$          & $0.3$       & $0.7$      & $^{+6.7}_{-2.2}$ & $^{+30}_{-15}$   & $^{+33}_{-19}$   \\
        $M_\xfourfive\,[\!\mev]$                & $0.4$         & $0.4$       & $1.4$      & $^{+2.7}_{-2.1}$ & $^{+0.6}_{-0.7}$ & $^{+3.2}_{-2.7}$ \\
        $\Gamma_\xfourfive\ \,[\!\mev]$         & $1.7$         & $0.2$       & $2.0$      & $^{+21}_{-8}$    & $^{+9.3}_{-3.1}$ & $^{+24}_{-9}$   \\
        $\sigma_\xfouronefour$                  & $31\%$        & $1.5\%$     & $0.9\%$    & $^{+1.6}_{-1.5}\%$ & $^{+6}_{-16}\%$    & $^{+32}_{-36}\%$   \\
        $\sigma_\xfourtwosevenfour$             & $19\%$        & $1.6\%$     & $0.8\%$    & $^{+7}_{-4}\%$     & $^{+6}_{-16}\%$    & $^{+22}_{-26}\%$   \\
        $\sigma_\xfourfive$                     & $3.2\%$       & $0.7\%$     & $0.2\%$    & $^{+2.5}_{-7.3}\%$ & $^{+16}_{-5}\%$    & $^{+17}_{-11}\%$   \\
        $\sigma_{\substack{\xfoursixeight \\
                           +\xfourseven}}$      & $5.3\%$       & $18\%$      & $7.3\%$    & $^{+2.5}_{-7.3}\%$ & $^{+6}_{-16}\%$    & $^{+25}_{-29}\%$   \\
        \hline
        \end{tabular}
    \label{tab:sys}
    \end{center}
\end{table}

\begin{table}[h]
    \centering
    \caption{\small Mass and width parameters obtained from the fit to the \jpsiphi invariant-mass distribution. The first uncertainty is statistical, while the second is systematic. The results of this analysis are compared to the ones observed in Ref.~\cite{LHCb-PAPER-2020-044}.}
    \begin{tabular}{c c c}
        \hline
        Parameter $[\mev]$          & Current analysis   & Ref.~\cite{LHCb-PAPER-2020-044} \\
        \hline
        $M_\xfourtwosevenfour$      & \Mfourtwosevenfour & $4294 \pm 4 ^{+3}_{-6}$         \\
        $\Gamma_\xfourtwosevenfour$ & \Wfourtwosevenfour &   $53 \pm 5 \pm 5$              \\
        \hline
        $M_\xfourfive$              & \Mfourfive         & $4474 \pm 3 \pm 3$              \\
        $\Gamma_\xfourfive$         & \Wfourfive         &   $77 \pm 6 ^{+10}_{-8}$        \\
        \hline
    \end{tabular}
    \label{tab:comparison}
\end{table}

In the fiducial region defined by $\mjpsiphi<6000\mev$ and muon and kaon candidates with $2<\eta<5$ and $\pt>200\mev$, and one muon candidate with $\pt>400\mev$, the resulting \jpsiphidecay production cross-section with no additional activity in the event is found to be
\begin{equation*}
    \begin{split}
        \sigma_{\jpsiphi} \times \BF(\jpsi\to&\mumu) \times \BF(\phi\to\Kp\Km) \\
                                            &= \Totalcs \mathrm{,}
    \end{split}
\end{equation*}
where the first uncertainty is statistical, the second is systematic and the third is due to the luminosity determination.
The individual production cross-sections for each of the observed resonant states are found to be
\begin{equation*}
    \begin{split}
        \sigma_{\xfouronefour}              \times \BF_{\mathrm{eff}}^{\xfouronefour} &= \xfouronefourcs \mathrm{,} \\
        \sigma_{\xfourtwosevenfour}         \times \BF_{\mathrm{eff}}^{\xfourtwosevenfour} &= \xfourtwosevenfourcs \mathrm{,} \\
        \sigma_{\xfourfive}                 \times \BF_{\mathrm{eff}}^{\xfourfive} &= \xfourfivecs \mathrm{,} \\
        \sigma_{\xfoursixeight+\xfourseven} \ \ \ \ & \\
                                            \times \BF_{\mathrm{eff}}^{\xfoursixeight+\xfourseven} &= \xfoursevencs \mathrm{,} \\
        \sigma_{\nr}                        \times \BF_{\mathrm{eff}}^{\nr} &= \NRcs \mathrm{,}
    \end{split}
\end{equation*}
where the first uncertainty is statistical, while the second is the systematic uncertainty combined with the luminosity determination uncertainty. Here, $\BF_{\mathrm{eff}}^{X} = \BF(X\to\jpsi\phi) \times \BF(\jpsi\to\mumu) \times \BF(\phi\to\Kp\Km)$, with $X$ denoting the five resonant states in the model, and $\BF_{\mathrm{eff}}^{\nr} = \BF(\jpsi\to\mumu) \times \BF(\phi\to\Kp\Km)$.

The statistical significance for the \xfouronefour resonance is estimated via Wilks' theorem~\cite{Wilks:1938dza} since its mass and width are fixed in the fit.
Even though Wilks' theorem applies strictly to cases where the resonance parameters are kept fixed, the analysis presented in this Letter follows the previous results obtained from $\Bu\to\jpsiphi\Kp$ decays~\cite{LHCb-PAPER-2020-044}.
Since the shape parameters of the \xfourtwosevenfour and \xfourfive resonances approximately agree with the expectations, the estimation of the look-elsewhere effect~\cite{Algeri_2016,Gross2010} is neglected and Wilks' theorem is applied in this case as well.
Furthermore, when computing the significance of one of them, the mass and width of the others are kept fixed.
The precision of the current analysis does not allow the \xfoursixeight and \xfourseven resonances to be resolved, therefore a joint significance is calculated.
For the individual significances, systematic uncertainties are taken into account using the procedure employed in Ref~\cite{LHCb-PAPER-2012-002}.
The significance for the resonances \xfouronefour, \xfourtwosevenfour and \xfourfive are found to be $2.3\,\sigma$, $4.1\,\sigma$ and $6.1\,\sigma$.
The joint \xfoursixeight+\xfourseven significance is found to be $1.8\,\sigma$.

In summary, 989 \jpsiphidecay candidates are observed in events without additional activity and the total fiducial cross-section is measured.
Several clear resonant structures are observed in the invariant-mass distribution of these \jpsiphi candidates, which is well-described by a model containing five resonant and one nonresonant components.
This is the first observation of \XtoJPsiPhi production in diffractive processes and therefore helps determine the underlying nature of exotic states.

\section*{Acknowledgements}
%
%
\noindent We express our gratitude to our colleagues in the CERN
accelerator departments for the excellent performance of the LHC. We
thank the technical and administrative staff at the LHCb
institutes.
We acknowledge support from CERN and from the national agencies:
CAPES, CNPq, FAPERJ and FINEP (Brazil); 
MOST and NSFC (China); 
CNRS/IN2P3 (France); 
BMBF, DFG and MPG (Germany); 
INFN (Italy); 
NWO (Netherlands); 
MNiSW and NCN (Poland); 
MCID/IFA (Romania); 
MICIU and AEI (Spain);
SNSF and SER (Switzerland); 
NASU (Ukraine); 
STFC (United Kingdom); 
DOE NP and NSF (USA).
We acknowledge the computing resources that are provided by CERN, IN2P3
(France), KIT and DESY (Germany), INFN (Italy), SURF (Netherlands),
PIC (Spain), GridPP (United Kingdom), 
CSCS (Switzerland), IFIN-HH (Romania), CBPF (Brazil),
and Polish WLCG (Poland).
We are indebted to the communities behind the multiple open-source
software packages on which we depend.
Individual groups or members have received support from
ARC and ARDC (Australia);
Key Research Program of Frontier Sciences of CAS, CAS PIFI, CAS CCEPP, 
Fundamental Research Funds for the Central Universities, 
and Sci. \& Tech. Program of Guangzhou (China);
Minciencias (Colombia);
EPLANET, Marie Sk\l{}odowska-Curie Actions, ERC and NextGenerationEU (European Union);
A*MIDEX, ANR, IPhU and Labex P2IO, and R\'{e}gion Auvergne-Rh\^{o}ne-Alpes (France);
AvH Foundation (Germany);
ICSC (Italy); 
Severo Ochoa and Mar\'ia de Maeztu Units of Excellence, GVA, XuntaGal, GENCAT, InTalent-Inditex and Prog. ~Atracci\'on Talento CM (Spain);
SRC (Sweden);
the Leverhulme Trust, the Royal Society
 and UKRI (United Kingdom).

\newpage


\addcontentsline{toc}{section}{References}
\bibliographystyle{LHCb}
\bibliography{main,standard,LHCb-PAPER,LHCb-CONF,LHCb-DP,LHCb-TDR}

\newpage
\centerline
{\large\bf LHCb collaboration}
\begin
{flushleft}
\small
R.~Aaij$^{35}$\lhcborcid{0000-0003-0533-1952},
A.S.W.~Abdelmotteleb$^{54}$\lhcborcid{0000-0001-7905-0542},
C.~Abellan~Beteta$^{48}$,
F.~Abudin{\'e}n$^{54}$\lhcborcid{0000-0002-6737-3528},
T.~Ackernley$^{58}$\lhcborcid{0000-0002-5951-3498},
A. A. ~Adefisoye$^{66}$\lhcborcid{0000-0003-2448-1550},
B.~Adeva$^{44}$\lhcborcid{0000-0001-9756-3712},
M.~Adinolfi$^{52}$\lhcborcid{0000-0002-1326-1264},
P.~Adlarson$^{79}$\lhcborcid{0000-0001-6280-3851},
C.~Agapopoulou$^{46}$\lhcborcid{0000-0002-2368-0147}
% ...

{\footnotesize \it

% ...
$^{35}$Nikhef National Institute for Subatomic Physics, Amsterdam, Netherlands\\
% ...
$^{44}$Instituto Galego de F{\'\i}sica de Altas Enerx{\'\i}as (IGFAE), Universidade de Santiago de Compostela, Santiago de Compostela, Spain\\
% ...
$^{46}$European Organization for Nuclear Research (CERN), Geneva, Switzerland\\
% ...
$^{48}$Physik-Institut, Universit{\"a}t Z{\"u}rich, Z{\"u}rich, Switzerland\\
% ...
$^{52}$H.H. Wills Physics Laboratory, University of Bristol, Bristol, United Kingdom\\
% ...
$^{54}$Department of Physics, University of Warwick, Coventry, United Kingdom\\
% ...
$^{57}$School of Physics and Astronomy, University of Glasgow, Glasgow, United Kingdom\\
$^{58}$Oliver Lodge Laboratory, University of Liverpool, Liverpool, United Kingdom\\
% ...
$^{66}$Syracuse University, Syracuse, NY, United States\\
% ...
$^{79}$Department of Physics and Astronomy, Uppsala University, Uppsala, Sweden, associated to $^{57}$\\
% ...
}
\end{flushleft}

\end{document}